# Spontaneous decay of artificial atoms in a multi-qubit system

Ya. S. Greenberg[1], A. A. Shtygashev[1], and A. G. Moiseev[1]

[1]Novosibirsk State Technical University, K. Marx Ave. 20, Novosibirsk, 630073

E-mail: yakovgreenberg@yahoo.com

## Abstract

*yakovgreenberg@yahoo.com

We consider a one-dimensional chain of N equidistantly spaced noninteracting qubits embedded in an open waveguide. In the frame of single-excitation subspace, we systematically study the evolution of qubits' amplitudes if the only qubit in the chain was initially excited. We show that the temporal dynamics of qubits' amplitudes crucially depend on the value of kd, where k is the wave vector, d is a distance between neighbor qubits. If kd is equal to an integer of $\pi$, then the qubits are excited to a stationary level which scales as $N^{-1}$. We show that in this case, it is the dark states which prevent qubits from decaying to zero even though they do not contribute to the output spectrum of photon emission. For other values of kd the excitations of qubits have the form of damping oscillations, which represent the vacuum Rabi oscillations in a multi-qubit system. In this case, the output spectrum of photon radiation is defined by a subradiant state with the smallest width.

## I. Introduction

Superconducting qubits coupled to photons propagating in an open waveguide [1–5] allow for investigation of the fascinating world of quantum light-matter interactions in one dimension [6–8].

The interesting feature of these systems is that the photon-mediated interaction between qubits is of infinite range. This can give rise to the formation of the multiple super or subradiant collective states that can decay at a rate faster or slower than the rate $\Gamma$ of a single qubit alone [3]. The properties of collective states in multi-qubit 1D systems have been extensively studied [9,10]. It has been shown that the decay rates of the collective states, $\Gamma_\xi$ scales as $\xi^2/N^3$, where the collective states are ordered from subradiant state with the smallest $\Gamma_\xi$ ($\xi=1$) to superradiant state with $\Gamma_\xi=\Gamma$ ($\xi=N$) [11].

Another remarkable consequence of the multi-qubit system is that a chain of many equidistant qubits acts as a nearly perfect mirror for an incident field close to resonance [12,13].

As is known, the superconducting qubits can be technologically addressed and controlled individually [8]. Therefore, it is important to know the evolution of the probability amplitude of any qubit in a superconducting circuitry.

Here, we will discuss the decay dynamics of any qubit's amplitude $\beta_n(t)$ when a single qubit in the chain is initially excited. As is known, the decay of a specific qubit can be described by a set of characteristic complex frequencies defined by the poles of corresponding collective state eigenvectors. However, for many qubits (practically, for N>3) the calculation of $\beta_n(t)$ based on the superposition of collective states is not mathematically convenient. Instead, in this paper, we show that the qubits' amplitudes $\beta_n(t)$ can be numerically calculated from a set of linear differential equations which allow us to avoid the use of collective states. We show that the temporal dynamics of qubits' amplitudes crucially depend on the value of kd, where k is the wave vector, d is a distance between neighbor qubits. If kd is equal to integer of $\pi$, the qubits are excited to a stationary level which scales as $N^{-1}$. We show that in this case, these are the dark states which prevent qubits from decaying to zero. For other values of kd, the excitations of qubits have oscillatory behavior and are gradually damped out to zero.

The paper is structured as follows. In Sec. II, we begin by introducing a Jaynes-Cummings Hamiltonian for atom-light interactions. A trial wave function is written in a single-photon excitation subspace. In Sec. III we present a non-Hermitian Hamiltonian which is obtained after the elimination of photon variables. We describe the collective states which are eigenvectors of non- Hermitian Hamiltonian and show that the application of collective states for the calculation of qubits' amplitudes is not mathematically convenient. In Sec. IV we obtain for the qubits' amplitudes $\beta_n(t)$ a set of linear differential equations which allow for a direct numerical simulations. In Sec. V, we provide a comprehensive analysis of the temporal dynamics of the qubits' amplitudes. We show that for kd=n$\pi$ the excited qubits' amplitudes become «frozen» at the level of 1/N which is explained by N−1 dark states which prevent qubits from decaying to zero. For the values of kd which are not integer of $\pi$ the qubits' amplitudes gradually damp out to zero. The probability amplitude of the photon emission is studied in Sec. VI. We find, that for kd=(2n+1)$\pi$ the evolution of the photon amplitude consists of many clearly seen steps. These steps can be attributed to the interrelation between temporal behaviors of the different qubits' amplitudes. In Sec. VII, we calculate a spectral density of photon radiation from a linear chain of N qubits. We show that for kd=nπ the dark states do not contribute to output radiation. However, if kd is not equal to an integer of π, a structure of spectral line is defined by the deep subradiant states. The main results of the paper are summarized in the concluding Section VIII.

## II. Formulation of the problem

We consider a linear chain of N equally spaced qubits which are coupled to photon field in an open waveguide (see Fig. 1).

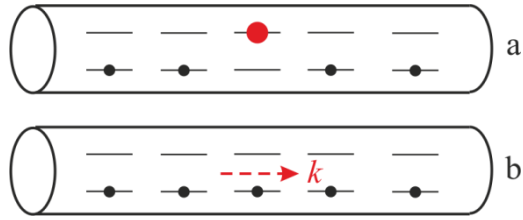

Fig. 1. Schematic illustration of a single-excitation subspace for a five-qubit chain in an open waveguide. (a) A single qubit is excited, N−1 qubits are in the ground state. (b) N qubits are in the ground state and a single photon propagates in the waveguide.

A distance between neighbor qubits is equal to $d$. The Hilbert space of every qubit consists of the excited state $|e\rangle$ and the ground state $|g\rangle$. The Hamiltonian which accounts for the interaction between qubits and the electromagnetic field is (we use units where $\hbar = 1$ throughout the paper):

$$H = H_0 + \sum_k \omega_k a_k^+ a_k + \sum_{n=1}^{N}\sum_k \left( \left( g_k^{(n)} e^{-ikx_n} \sigma_-^{(n)} \right) a_k^+ + h.c. \right) \quad (1)$$

where $H_0$- is Hamiltonian of bare qubits.

$$H_0 = \frac{1}{2}\sum_{n=1}^{N}\left(1+\sigma_z^{(n)}\right)\Omega_n \quad (2)$$

The quantity $g_k^{(n)}$ in (1) is the coupling between n-th qubit and the photon field in a waveguide.

Below we consider a single-excitation subspace with either a single photon is in a waveguide and all qubits are in the ground state, Fig. 1b, or there are no photons in a waveguide with the only n-th qubit in the chain being excited, Fig. 1a. Therefore, we have limited Hilbert space to the following states:

$$|n,0_k\rangle = |n\rangle \otimes |0_k\rangle = |g_1, g_2, \ldots g_{n-1}, e_n, g_{n+1}, \ldots g_N\rangle \otimes |0_k\rangle$$
$$|G,1_k\rangle = |G\rangle \otimes |1_k\rangle = |g_1, g_2, \ldots g_{N-1}, g_N\rangle \otimes |1_k\rangle \quad (3)$$

The Hamiltonian (1) preserves the number of excitations (number of excited qubits + number of photons). In our case the number of excitations is equal to one (see Fig. 1). Therefore, the system will remain within a single-excitation subspace at any instant of time. The wave function of an arbitrary single-excitation state can then be written in the form:

$$|\Psi\rangle = \sum_{n=1}^{N} \beta_n(t) e^{-i\Omega_n t} |n,0_k\rangle + \sum_k \gamma_k(t) e^{-i\omega_k t} |G,1_k\rangle, \quad (4)$$

where $\beta_n(t)$ is the amplitude of n-th qubit, $\gamma_k(t)$ is a single-photon amplitude which is related to a spectral density of spontaneous emission.

$$S(\omega_k, t) = |\gamma_k(t)|^2 \quad (5)$$

The function (4) is normalized to unity:

$$\sum_{n=1}^{N} |\beta_n(t)|^2 + \sum_k |\gamma_k(t)|^2 = 1 \quad (6)$$

Our goal is to find the evolution of the amplitudes $\beta_n(t)$ for any qubit in the chain with the initial conditions when the only $n_0$-th qubit in a chain is excited at *t=0*:

$$\beta_{n_0}(0) = 1;$$
$$\beta_n(0) = 0; n \neq n_0 \quad (7)$$

### III. Effective Hamiltonian and collective states

In what follows, we assume all qubits are identical concerning their excitation frequency $\Omega$ and the rate $\Gamma$ of spontaneous emission of individual qubits into the waveguide. The elimination of photon variables results in a non-Hermitian effective Hamiltonian, which in the Markovian approximation accounts for the photon-mediated interaction between qubits [12, 14]

$$H_{eff} = -i\frac{\Gamma}{2} \sum_{m,n=1}^{N} e^{ik|x_m - x_n|} \sigma_m^+ \sigma_n \quad (8)$$

Where $k=\Omega/v_g$, $v_g$ is the group velocity of electromagnetic wave in a waveguide, $x_n$ is the position of n-th qubit, $\sigma_n^+, \sigma_n$ are raising and lowering spin operators for n-th qubit. The rate of spontaneous emission $\Gamma$ of an individual qubit is defined by the Fermi golden rule:

$$\Gamma = 2\pi \sum_k |g_k|^2 \delta(\omega_k - \Omega) \tag{9}$$

It follows from (8) that the photon-mediated interaction between qubits in such a system results in coherent $J_{mn} = \Gamma \sin(k|x_m - x_n|)/2$ and dissipative rates $\Gamma_{mn} = \Gamma \cos(k|x_m - x_n|)$. The coherent rate shifts the positions of the qubits' resonances, while the dissipative rate gives rise to the additional spontaneous emission into the waveguide mode. Unlike real atoms with short-range dipole-dipole interaction, here a coherent interaction $J_{mn}$ is a long-range one: every qubit is sensitive to its distant neighbor. In a single-excitation subspace Hamiltonian (8) has N collective eigenfunctions which are obtained from the Schrodinger equation $H_{eff}\Psi = E\Psi$.

$$|\Psi_i(t)\rangle = e^{-i\bar{E}_i t} \sum_{n=1}^{N} \alpha_n^{(i)} |n\rangle; (i = 1, 2 \ldots N) \tag{10}$$

where $\bar{E}_i$ is a complex energy

$$\bar{E}_i = E_i - i\frac{\Gamma_i}{2} \tag{11}$$

The quantities $E_i$ and $\Gamma_i$ depend on the system parameters $\Gamma$, k, $x_n$. The wave vectors with $\Gamma_i<\Gamma$ are called subradiant states, those with $\Gamma_i>\Gamma$ are called superradiant states.

Complex energies can be found by equating to zero the determinant of the matrix.

$$\left(E + i\frac{\Gamma}{2}\right)\delta_{mn} + i\frac{\Gamma}{2} e^{ik|x_m - x_n|}(1 - \delta_{mn}) \tag{12}$$

An important sum rule

$$\sum_{i=1}^{N} \Gamma_i = N\Gamma \tag{13}$$

states that there are no other losses in the system other than the coherent photon emission into a waveguide.

Since Hamiltonian (8) is non-Hermitian, the set of eigenfunctions (10) is neither normalized nor orthonormal. It is known that a correct calculation of the coefficients $\alpha_n^{(i)}$ in (10) requires a bi-orthogonal set of eigenfunctions $|\bar{\Psi}_i(t)\rangle$ which are a solution of the Schrodinger equation for $H_{eff}^\dagger$. In our case $H_{eff}^\dagger = H_{eff}^*$ with the consequence that the complex conjugate of an eigenstate $|\Psi_i(t)\rangle$ of H$_{eff}$ is an eigenstate of $H_{eff}^\dagger$. Therefore, the conditions for normalization and orthonormality between eigenfunctions of these two sets lead to the following equations for the coefficients $\alpha_n^{(i)}$ [15, 16]:

$$\langle \bar{\Psi}_i(t) | \Psi_i(t) \rangle = \sum_{n=1}^{N} \left( \alpha_n^{(i)} \right)^2 = 1 \tag{14a}$$

$$\langle \bar{\Psi}_i(t) | \Psi_j(t) \rangle = e^{i(E_i - E_j)t} \sum_{n=1}^{N} \alpha_n^{(i)} \alpha_n^{(j)} = 0 \tag{14b}$$

We can express the qubits' amplitudes β$_n$(t) in terms of the coefficients $\alpha_n^{(i)}$. First, we write the dynamic wave function as a decomposition over the collective states:

$$|\Psi(t)\rangle = \sum_{i=1}^{N} A_i |\Psi_i(t)\rangle = \sum_{i,n=1}^{N} A_i e^{-i\bar{E}_i t} \alpha_n^{(i)} |n\rangle \tag{15}$$

From (15) we see that

$$\beta_n(t) = \sum_{i=1}^{N} e^{-i\bar{E}_i t} A_i \alpha_n^{(i)} \tag{16}$$

with the initial conditions

$$\beta_{n_0}(0) = \sum_{i=1}^{N} A_i \alpha_p^{(i)} = 1$$

$$\beta_n(0) = \sum_{i=1}^{N-1} A_i \alpha_n^{(i)} = 0; n \neq n_0 \tag{17}$$

This procedure looks elegant but it is not convenient for computer simulations if the number of the qubits is large. It consists of several steps. The first step requires the calculation of N complex energies $\bar{E}_i$ from determinant (12). The second step is the calculations of $\alpha_n^{(i)}$ using non linear conditions (14a) and (14b). And finally, the third step requires finding the solution of a system of N linear algebraic equations (17). Every one of these three steps is not simple from a mathematical point of view. As an alternative approach, we obtain below a set of the linear differential equations for the qubits' amplitudes β$_n$(t), which allow us to find the quantities β$_n$(t) by direct computer simulations of these equations.

## IV. The equations for qubits' amplitudes

Here, we express the wave function for Hamiltonian (8) in terms of a superposition of the single excited states

$$\Psi(t) = \sum_{n=1}^{N} \beta_n(t) |n\rangle \tag{18}$$

Even though the wave functions (4) and (18) are different, the qubits' amplitudes $\beta_n(t)$ in these expressions are the same.

The equations for $\beta_n(t)$ are derived from the time-dependent Schrodinger equation $id\Psi/dt = H_{eff}\Psi$. We obtain

$$\frac{d\beta_n}{dt} = -\frac{\Gamma}{2}\beta_n(t) - \frac{\Gamma}{2}\sum_{m \neq n}^{N-1} \beta_m(t) e^{ik|x_m - x_n|} \tag{19a}$$

If qubits are equidistantly spaced by the distance *d*, the equation (19a) can be rewritten as follows:

$$\frac{d\beta_n}{dt} = -\frac{\Gamma}{2}\beta_n(t) - \frac{\Gamma}{2}\sum_{m \neq n}^{N-1} \beta_m(t) e^{ikd|m-n|} \tag{19b}$$

where the qubits are ordered from left to right, and n is the number of a qubit in the chain.

The solution of equations (19a) has the form:

$$\beta_n(t) = \sum_{i=1}^{N} b_i^{(n)} e^{\lambda_i t} \tag{20}$$

where coefficients $b_i^{(n)}$ are defined by the initial conditions (7). The quantities λ$_i$ are characteristic roots, which can be found by equating to zero the determinant of the matrix

$$\left(\lambda_i + \frac{\Gamma}{2}\right)\delta_{mn} + \frac{\Gamma}{2} e^{ikd|m-n|}(1 - \delta_{mn}) \tag{21}$$

A comparison of (21) with (12) gives a simple relation between quantities $\lambda_n$ and complex energies, $\bar{E}_n$: $\bar{E}_n = i\lambda_n$.

The number of the quantities $\lambda_n$ is exactly equal to the number of qubits N. In principle, the evolution of qubits' amplitudes is given by expression (20). However, only a limited number of simple cases can be analytically obtained in the form of (20). In general, the qubits' amplitudes can be only obtained by numerical simulation of equations (19). The equation (19b) with initial conditions (7) is a starting point for our subsequent calculations.

## V. The evolution of qubits' amplitudes

We assume that initially, the $n_0$-th qubit is excited, while all other qubits are in the ground state. First, we consider several simple cases where the analytic solutions can be obtained.

**1. kd=2πn, where n is integer.**

It follows from the symmetry of equations (19b) that the evolution of all qubits, except for that of excited one, is the same. That is, all $\beta_n(t), n \neq n_0$ are equal. Therefore, N equations (19b) can be reduced to two equations:

$$\frac{d\beta_n}{dt} = -\frac{\Gamma}{2}(N-1)\beta_n(t) - \frac{\Gamma}{2}\beta_{n_0}(t)$$
$$\frac{d\beta_{n_0}}{dt} = -\frac{\Gamma}{2}\beta_{n_0}(t) - \frac{\Gamma}{2}(N-1)\beta_n(t)$$

(22)

The characteristic roots of this system are as follows:

$$\lambda_1 = 0; \quad \lambda_2 = -\frac{\Gamma N}{2}$$

(23)

These roots are the signature of a superradiant decay. Here for N qubits, we have a single decaying state with the energy $E_r = \Omega - iN\Gamma/2$ and $N-1$ dark states which do not interact with the photon field.

For the solution of (22) we obtain:

$$\beta_{n_0}(t) = \frac{N-1}{N} + \frac{1}{N}e^{-\frac{\Gamma}{2}Nt}$$
$$\beta_n(t) = -\frac{1}{N} + \frac{1}{N}e^{-\frac{\Gamma}{2}Nt}$$

(24)

It is seen from (24) that the decay rate is equal to $N\Gamma$. As the qubit number increases, the change of the qubits' amplitudes scales as $1/N$ at $t \to \infty$.

From condition (6) we can find the full probability of photon emission from the N-qubit system.

$$P_{ph}(t) \equiv \sum_k |\gamma_k(t)|^2 = \frac{1}{N}\left(1-e^{-\Gamma Nt}\right) \qquad (25)$$

Therefore, as N is increased the qubits' amplitudes are weakly changed, and the radiation is mostly blocked within a system.

**2. $kd=(2n+1)\pi$., where n is an integer.**

It should be expected that in this case the behavior of the qubits is divided into two groups, in each of which the qubits have the same amplitudes. One group includes qubits whose distance from an excited qubit is equal to an odd number of d values: d, 3d, 5d ... The distance of another group of qubits from an excited one is equal to an even number of d: 2d, 4d, 6d ... We denote by $\beta^{(1)}$ and $N_1$ the amplitude and the number of qubits in the first group, respectively, and by $\beta^{(2)}$ and $N_2$ the same quantities in the second group. Obviously, $N_1 + N_2 = N-1$.

Let us first write down the equation for the $n_0$-th excited qubit. It should be borne in mind that due to a phase factor in the sum of (19b), all the amplitudes belonging to the first group are taken with a negative sign, while the amplitudes of the second group are taken with a positive sign.

From (19b) we obtain for excited qubit:

$$\frac{d\beta_{n_0}}{dt} = -\frac{\Gamma}{2}\beta_{n_0}(t) + \frac{\Gamma}{2}N_1\beta^{(1)} - \frac{\Gamma}{2}N_2\beta^{(2)} \qquad (26)$$

For any not excited n-th qubit the equation (19b) is of the form (here we count qubits from the excited one, therefore, $m-n=(m-n_0)-(n-n_0)$)

$$\begin{aligned}\frac{d\beta_n}{dt} &= -\frac{\Gamma}{2}\beta_n(t) - \frac{\Gamma}{2}\sum_{m\neq n}^{N-1}\beta_m(t)e^{ikd|(m-n_0)-(n-n_0)|} \\ &= -\frac{\Gamma}{2}\beta_n(t) - \frac{\Gamma}{2}\beta_{n_0}(t)e^{ikd|(n-n_0)|} - \frac{\Gamma}{2}\sum_{m\neq n,n_0}^{N-2}\beta_m(t)e^{ikd|(m-n_0)-(n-n_0)|}\end{aligned} \qquad (27a)$$

Next, we split the last sum in (27a) into two parts. The first part consists of odd values of m–n₀, while the second part consists of even values of m–n₀. Then we obtain from (27a):

$$\frac{d\beta_n}{dt} = -\frac{\Gamma}{2}\beta_n(t) - \frac{\Gamma}{2}\beta_{n_0}(t)e^{ikd|(n-n_0)|} - \frac{\Gamma}{2}\sum_{m\neq n,n_0}^{N-2}\beta_m(t)e^{ikd|(m-n_0)-(n-n_0)|}$$

$$= -\frac{\Gamma}{2}\beta_n(t) - \frac{\Gamma}{2}\beta_{n_0}(t)e^{ikd|(n-n_0)|} \quad (27б)$$

$$-\frac{\Gamma}{2}\sum_{p=1}^{N_1}\beta_{2p-1+n_0}(t)e^{ikd|(2p-1)-(n-n_0)|} - \frac{\Gamma}{2}\sum_{p=1}^{N_2}\beta_{2p+n_0}(t)e^{ikd|2p-(n-n_0)|}$$

From (27b) we can write the equations for the amplitudes $\beta^{(1)}$ and $\beta^{(2)}$, bearing in mind that for $\beta^{(1)}$ the quantity n–n₀ is an odd number, while for $\beta^{(2)}$ the quantity n–n₀ is an even number.

$$\frac{d\beta^{(1)}}{dt} = -\frac{\Gamma}{2}N_1\beta^{(1)}(t) + \frac{\Gamma}{2}N_2\beta^{(2)}(t) + \frac{\Gamma}{2}\beta_{n_0}(t) \quad (28a)$$

$$\frac{d\beta^{(2)}}{dt} = -\frac{\Gamma}{2}N_2\beta^{(2)}(t) + \frac{\Gamma}{2}N_1\beta^{(1)}(t) - \frac{\Gamma}{2}\beta_{n_0}(t) \quad (28b)$$

Summing up these equations we obtain $\beta^{(1)}(t) + \beta^{(2)}(t) = const$ for any time. Since initially all qubits' amplitudes except for that of the excited one are zero, then we have at any time $\beta^{(1)}(t) + \beta^{(2)}(t) = 0$. It allows for the reduction of three equations (26), (28a), and (28b) to the system of two coupled equations.

$$\frac{d\beta_i}{dt} = -\frac{\Gamma}{2}\beta_{n_0}(t) + \frac{\Gamma}{2}(N-1)\beta^{(1)} \quad (29a)$$

$$\frac{d\beta^{(1)}}{dt} = \frac{\Gamma}{2}\beta_{n_0}(t) - \frac{\Gamma}{2}(N-1)\beta^{(1)} \quad (29б)$$

Formally, these equations are identical to equations (22). The characteristic roots of this system coincide with (23). Therefore, the solution for qubits' amplitudes for this case is as follows:

$$\beta_{n_0}(t) = \frac{N-1}{N} + \frac{1}{N}e^{-\frac{\Gamma}{2}Nt}$$

$$\beta^{(1)}(t) = \frac{1}{N} - \frac{1}{N}e^{-\frac{\Gamma}{2}Nt} \quad (30)$$

$$\beta^{(2)}(t) = -\frac{1}{N} + \frac{1}{N}e^{-\frac{\Gamma}{2}Nt}$$

The full probability of photon emission for this case coincides with the expression (25).

It worth noting that the solutions (24), (25), and (30) are valid if any qubit in the chain is initially excited. Below we show the results of direct simulations of equations (19b) for a five-qubit system.

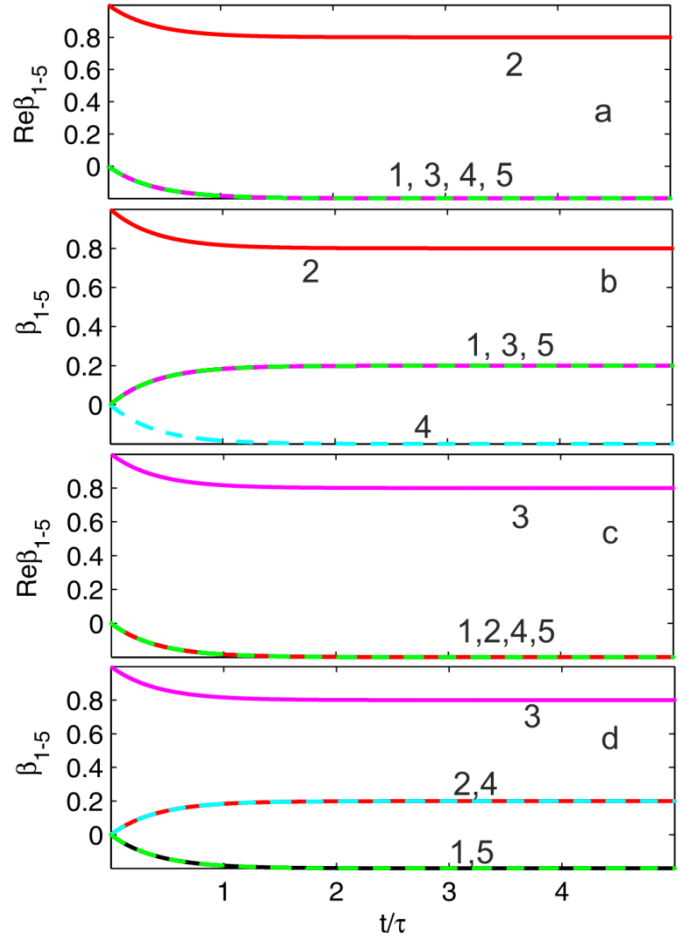

Fig. 2. The evolution of qubits' amplitudes for a five-qubit system. $\tau=1/\Gamma$. (a) The second qubit is initially excited. $kd=2\pi$. (b) The second qubit is initially excited. $kd=\pi$. (c) The third, a central, qubit is initially excited. $kd=2\pi$. (d) The third, a central, qubit is initially excited. $kd=\pi$.

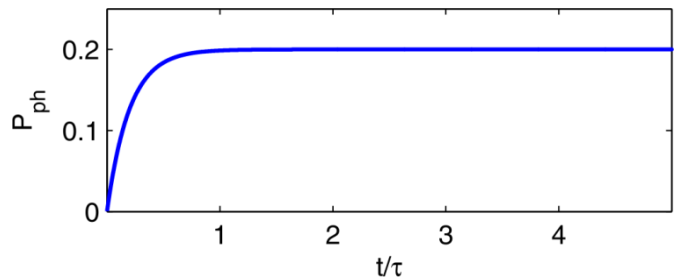

Fig. 3. The probability $P_{ph}(t)$ for the photon to be emitted. Five-qubit system, $kd=\pi, 2\pi$. $\tau=1/\Gamma$.

Obviously, these figures confirm the validity of analytical results obtained in (24), (25), and (30).

Therefore, for kd=nπ the characteristic time for the evolution of the qubits' amplitudes is on the order of 1/NΓ. It takes this time for a qubit amplitude to reach the level of 1/N. We may say that as the number N of qubits increases, the qubits' amplitudes become more «frozen». As is seen from (20) this frozenness is solely due to the dark states ($\lambda_n$=0), which prevent the qubits' amplitudes from decaying to zero.

The frozenness can be lifted if kd is not equal to an integer number of π. Then all qubits are damped to zero with the rate being determined by the root $\lambda_n$ with the smallest real part.

Below, for convenience, we consider the chain with an odd number of qubits. We assume for definiteness that a central qubit is initially excited. It is clear from the symmetry of the system that the qubits' which are located equally on both sides from the central qubit have the same amplitudes. Therefore, it allows for a reduction of N equations (19b) to (N+!)/2 equations. As an example, we consider a five-qubit system with the central qubit being initially excited, $\beta_3(0)$=1. For kd=(2n+1)π/2 the equations for qubits' amplitudes are as follows:

$$\frac{d\beta_1}{dt} = -\frac{\Gamma}{2}\beta_1 - i\frac{\Gamma}{2}\beta_2 + \frac{\Gamma}{2}\beta_3 + i\frac{\Gamma}{2}\beta_4 - \frac{\Gamma}{2}\beta_5$$

$$\frac{d\beta_2}{dt} = -i\frac{\Gamma}{2}\beta_1 - \frac{\Gamma}{2}\beta_2 - i\frac{\Gamma}{2}\beta_3 + \frac{\Gamma}{2}\beta_4 + i\frac{\Gamma}{2}\beta_5$$

$$\frac{d\beta_3}{dt} = \frac{\Gamma}{2}\beta_1 - i\frac{\Gamma}{2}\beta_2 - \frac{\Gamma}{2}\beta_3 - i\frac{\Gamma}{2}\beta_4 + \frac{\Gamma}{2}\beta_5 \qquad (31)$$

$$\frac{d\beta_4}{dt} = i\frac{\Gamma}{2}\beta_1 + \frac{\Gamma}{2}\beta_2 - i\frac{\Gamma}{2}\beta_3 - \frac{\Gamma}{2}\beta_4 - i\frac{\Gamma}{2}\beta_5$$

$$\frac{d\beta_5}{dt} = -\frac{\Gamma}{2}\beta_1 + i\frac{\Gamma}{2}\beta_2 + \frac{\Gamma}{2}\beta_3 - i\frac{\Gamma}{2}\beta_4 - \frac{\Gamma}{2}\beta_5$$

The numerical calculation provides five roots for this system: $\lambda_1$= −0.05Γ−i0.59Γ, $\lambda_2$=−0.05Γ+i0.59Γ, $\lambda_3$= −0.50Γ−i0.86Γ, $\lambda_4$= −0.50Γ+i0.86Γ, $\lambda_5$= −1.40Γ. However, due to the symmetry of the system, $\beta_1(t) = \beta_5(t); \beta_2(t) = \beta_4(t)$. Therefore, five equations (31) can be reduced to three equations.

$$\frac{d\beta_1}{dt} = -\Gamma\beta_1 + \frac{\Gamma}{2}\beta_3$$
$$\frac{d\beta_2}{dt} = -i\frac{\Gamma}{2}\beta_3 \qquad (32)$$
$$\frac{d\beta_3}{dt} = \Gamma\beta_1 - i\Gamma\beta_2 - \frac{\Gamma}{2}\beta_3$$

The characteristic roots of this system can be found from cubic equation:

$$\lambda(\lambda+\Gamma)\left(\lambda+\frac{\Gamma}{2}\right)+\frac{\Gamma^3}{2} = 0 \qquad (33)$$

The solution of this equation provides three characteristic roots $\lambda_1 = -0.05\Gamma - i0.59\Gamma$, $\lambda_2 = -0.05\Gamma + i0.59\Gamma$, $\lambda_5 = -1.40\Gamma$, which belong to the set of the roots for the five-equation system (31). It means that the decay of the central qubit excites only three collective states, associated with these three roots. The evolution of qubits for this case is shown in Fig. 4.

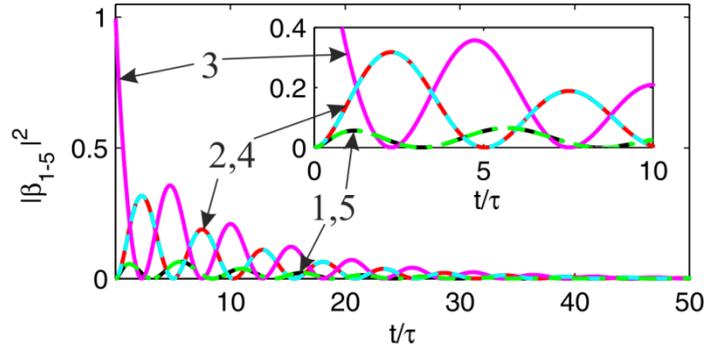

Fig. 4. Time dependence of qubits' amplitudes for a five-qubit system with a central qubit being excited. kd=π/2, τ=1/Γ. The numbers on the panel denote the qubit number ordered from left to right. For clarity, the evolution of qubits' amplitudes on a smaller time scale is shown in the insert.

As is seen from Fig. 4 the qubits' amplitudes are damped out as a distance between the unexcited qubit and the excited one is increased. This property is also illustrated in Fig. 5 for the nine-qubit system.

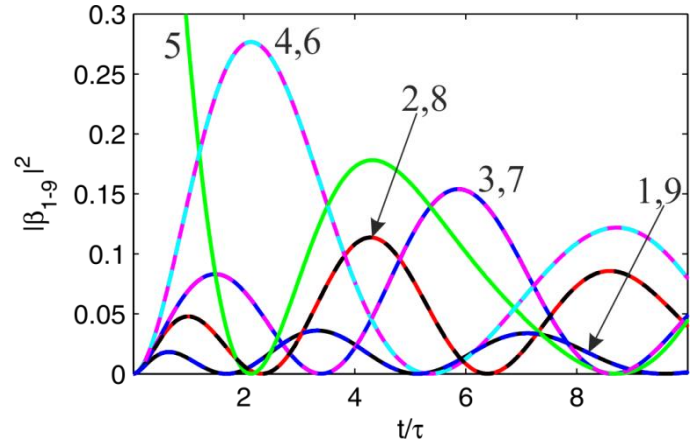

Fig.5. The evolution of qubits' amplitudes for nine-qubit system. The central, fifth qubit is initially excited. $kd=\pi/2$, $\tau=1/\Gamma$. The numbers in the figure denote the qubit numbers in the chain ordered from left to right.

As is known, a single qubit totally reflects incident photon if its frequency exactly equals that of a qubit [1,17]. However, for an incident field near the resonance a multi-qubit system exhibits a wide forbidden frequency gap for a transmittance of light [12,13]. On the other hand, for any qubit inside a waveguide, the waveguide acts as a high-quality resonator. As a consequence, the oscillations of qubits' amplitudes in Figs. 4, 5 are a clear signature of vacuum Rabi oscillations in a multi-qubit system.

For other values of kd, the evolution of qubits' amplitudes may be rather complicated (see Fig. 6.). It is mainly defined by the distribution of the roots of $\lambda_n$, which are shown in Fig. 7.

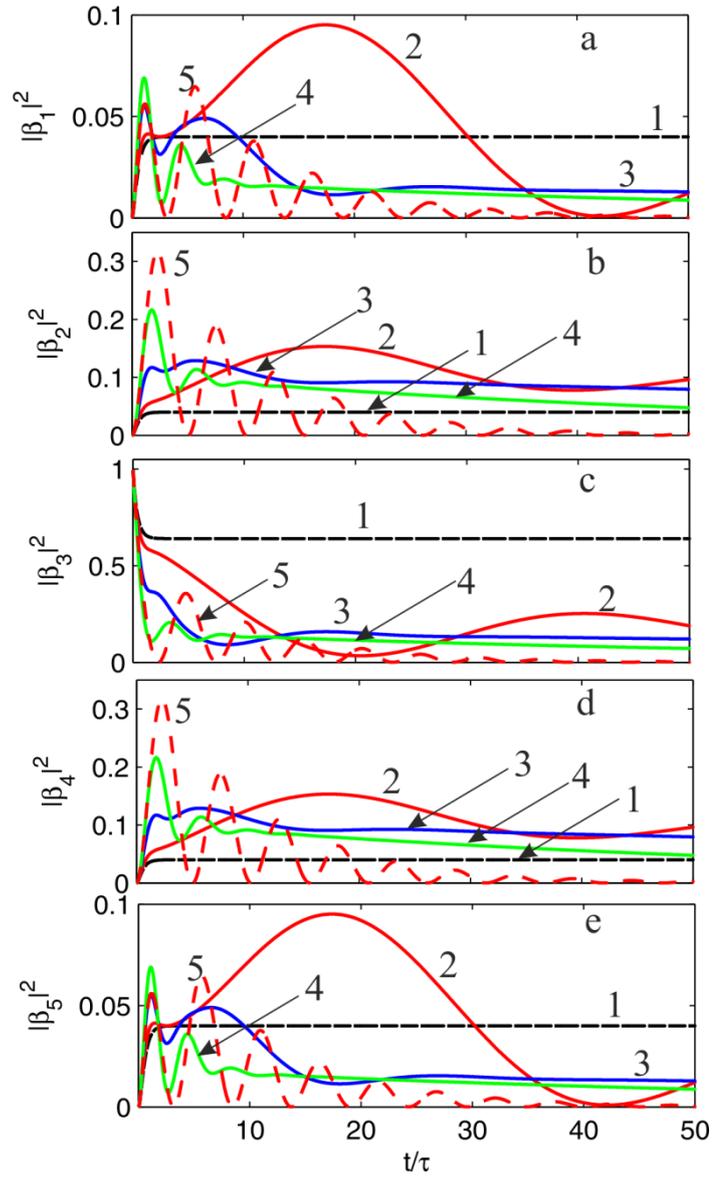

Fig. 6. The evolution of qubits' probabilities for a five-qubit system with a third, central, qubit being excited. The numbers on the panels denote the different values of the quantity kd. (1) kd=2π-black dashed line; (2) kd=2.1π-red solid line; (3) kd=2.2π-blue solid line; (4) kd=2.3π-green solid line; (5) kd=2.4π-red dashed line. τ=1/Γ.

From Fig. 6 it is clearly seen that the evolution of qubits that are equally spaced from the central qubit is similar.

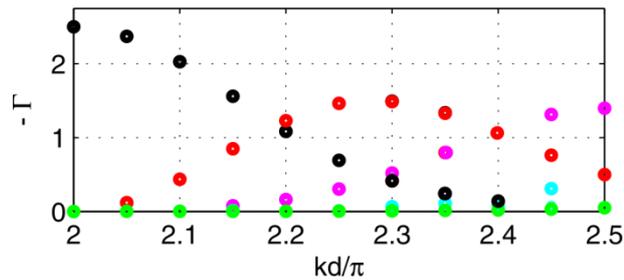

Fig. 7. Distribution of Re($\lambda_n$) for a five-qubit system.

The distribution of real parts of $\lambda_n$ is shown in Fig.7. For any kd which is not equal to $n\pi$ there exist, in general, five different roots. Not all of them are seen in Fig. 7.

As the number of qubits in the chain increases, the amplitudes of qubit oscillations, in general, decrease. This behavior is shown in Fig. 8. The reason for this is that the energy of the excited qubit must be distributed over all other qubits in the chain.

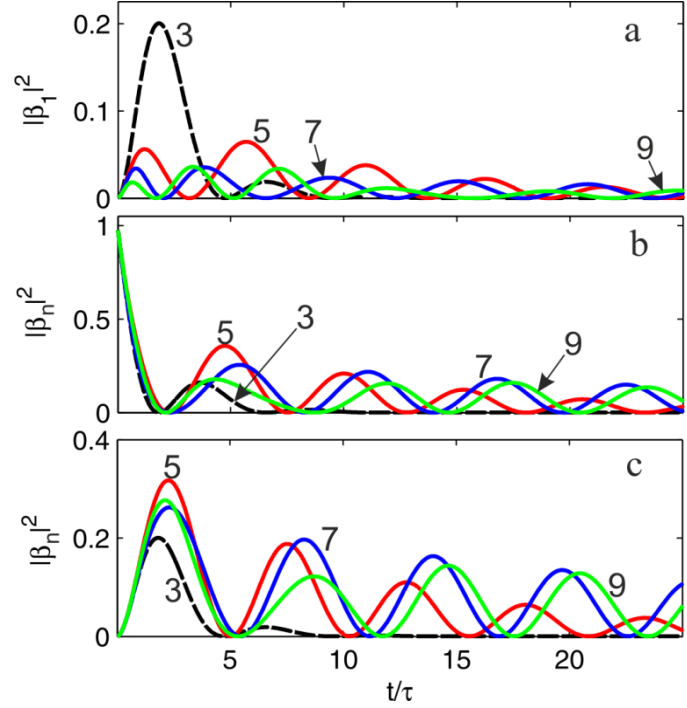

Fig.8. Dependence of qubits' amplitudes on N. $kd=\pi/2$. The numbers on the panels denote the number of qubits in the chain. In all cases a central qubit is initially excited. (a) The evolution of the first qubit in the chain. (b) The evolution of the central excited qubit. (c) The evolution of the qubit which is nearest to the central one. $\tau=1/\Gamma$.

## VI. The probability of the photon emission

Once we know the qubits' amplitudes $\beta_n(t)$, we may calculate the evolution of the full (integrated over all frequencies) probability of the photon emission, $P_{ph}(t) = \sum_k |\gamma_k(t)|^2$.

From (6) we obtain

$$P_{ph}(t) = 1 - \sum_{n=1}^{N} |\beta_n(t)|^2 \qquad (34)$$

We have found an interesting feature of this quantity if kd is equal to half-integer of $\pi$. The evolution of $P_{ph}(t)$ reveals several steps where, $dP_{ph}(t)/dt = 0$. These steps can be attributed to the interrelation between the temporal dynamics of the different amplitudes. As an example, we demonstrate this property in Fig. 9 for a three-qubit system.

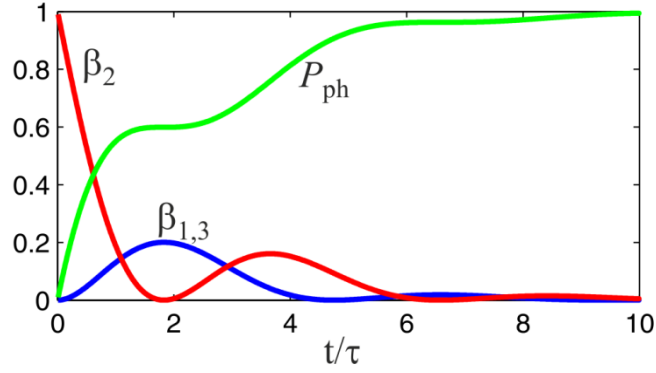

Fig. 9. Three-qubit system. The second qubit is initially excited. kd=$\pi$/2, $\tau$=1/$\Gamma$.

As is seen from this figure, the first step on $P_{ph}$ curve is in the vicinity of extremum points of qubits' amplitudes. However, for more qubits in the chain, the relation between a particular step and concrete qubits' amplitudes is not so evident. As an example, we show in Fig.10 several photon steps for a five-qubit system.

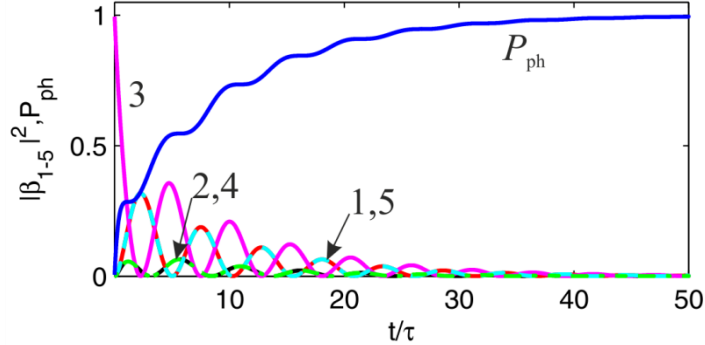

Fig. 10. The photon emission probability steps for the five-qubit system shown in Fig. 4. Central qubit is initially excited. kd=$\pi$/2, $\tau$=1/$\Gamma$.

## VII. Spectral density of photon radiation

The quantity $\gamma_k(t)$ in (4) allows for the calculation of a spectral density of spontaneous emission into a waveguide.

The application of the Schrodinger equation to wave function (4) leads to the following expression for $\gamma_k(t)$:

$$\gamma(\omega,t) = -i\sum_{n=1}^{N} g_k^{(n)} e^{-ikx_n} \int_0^t \beta_n(t') e^{i(\omega-\Omega_n)t'} dt' \quad (35a)$$

Using the expression (12) for $\beta_n(t)$ we obtain

$$\gamma(\omega,t) = \sum_{n=1}^{N} \sum_j g_k^{(n)} b_j^{(n)} e^{-ikx_n} \frac{1 - e^{i(\omega-\Omega_n - i\lambda_j)t}}{\omega - \Omega_n - i\lambda_j} \quad (35b)$$

This is a rather general expression. As the time proceeds only the term with the smallest real part of $\lambda_j$ survives in (35b). However, the concise analytical results can be obtained only for several simple cases.

Consider $kd=2n\pi$ with the identical qubits. Substitution of $\beta_n(t)$ from (24) in (35a) allows us to proceed with simple calculations.

.

$$\gamma_k(t) = -ig_k e^{ikx_i} \left[ \frac{N-1}{N} \int_0^t e^{i(\omega-\Omega)t'} dt' + \frac{1}{N} \frac{e^{i\left(\omega-\Omega+i\frac{\Gamma N}{2}\right)t} - 1}{i\left(\omega-\Omega+i\frac{\Gamma N}{2}\right)} \right]$$

$$-ig_k \sum_{j\neq i}^{N-1} e^{-ikx_j} \left[ -\frac{1}{N} \int_0^t e^{i(\omega-\Omega)t'} dt' + \frac{1}{N} \frac{e^{i\left(\omega-\Omega+i\frac{\Gamma N}{2}\right)t} - 1}{i\left(\omega-\Omega+i\frac{\Gamma N}{2}\right)} \right] \quad (36)$$

If the qubits are equidistant with a distance $d$ between neighbors, then a coordinate $x_n$ of the n-th qubit can be expressed in terms of coordinate $x_1$ of the first qubit: $x_n = x_1 + (n-1)d$. For $x_1=0$ and $k=\Omega/v_g$ we obtain from (36) for $kd=2\pi n$:

$$\gamma_k(t) = -g_k \frac{e^{i\left(\omega-\Omega+i\frac{\Gamma N}{2}\right)t} - 1}{\left(\omega-\Omega+i\frac{\Gamma N}{2}\right)} \xrightarrow{t\to\infty} \frac{g_k}{\left(\omega-\Omega+i\frac{\Gamma N}{2}\right)} \quad (37)$$

Therefore, in this case, a spectral density of spontaneous emission has a Lorentian form with a full width at half the height of the resonance line being equal to $N\Gamma$.

Two important things concerning the expression (37) are worth to be mentioned. First, a wave vector *k* in (35a) and (35b) is related to a running frequency ω: $k = \omega / v_g$. Our assumption $k = \Omega / v_g$ means that the expression (37) is only valid in the near resonance region. Second, the time-independent terms in (24) cancel each other in (36). A physical reason for this is that these terms are related to the dark states ($\lambda_n$=0) which do not interact with a photon field and, therefore, cannot contribute to the photon emission.

Similar calculations for kd=(2n+1)π with the qubits' amplitudes (30) also give the expression (37) for the spectral density of spontaneous emission.

For the values of kd, which are different from integer of π, the spectral density can be calculated only numerically.

Below we show the spectral density for three and five-qubit systems obtained by numerical calculations. In both cases, the central qubit is initially excited and kd=π/2.

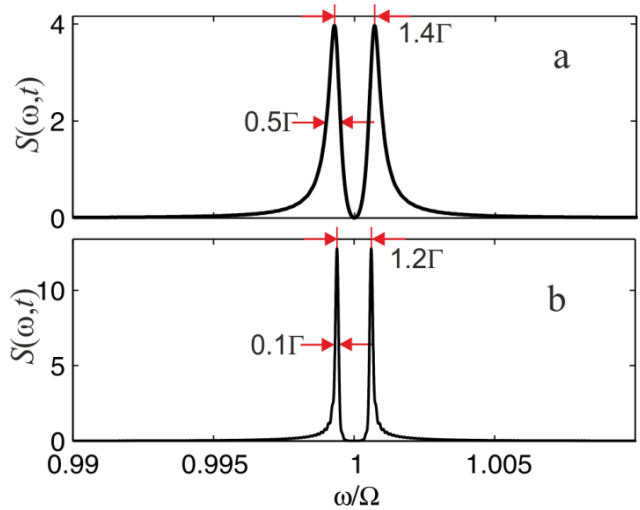

Fig. 11. Spectral density for three (a) and five (b) -qubit systems. kd=π/2, t=100/Γ. The spectrum in (a) is defined by the roots $\lambda_1 = -(0.25 + i \cdot 0.66)\Gamma$, $\lambda_2 = -(0.25 - i \cdot 0.66)\Gamma$. The spectrum in (b) is defined by the roots $\lambda_1 = -(0.05 + i \cdot 0.59)\Gamma$, $\lambda_2 = -(0.05 - i \cdot 0.59)\Gamma$.

The spectra in Fig. 11 are calculated for t=100/Γ, where only the roots $\lambda_n$ with the smallest real parts survive. It is known, that with the increase of N the real parts of the roots of deep subradiant states ($\Gamma_i$<<Γ) scale as $N^{-3}$ [11]. Therefore, the same scaling law should be observed for the widths of the spectral lines.

A spectrum of photon emission can be drastically changed if we take other values of kd or excite another qubit. Several spectra for the five-qubit system are shown in Fig. 12.

The form of the spectral line depends mainly on kd since this value defines the distribution of characteristic roots $\lambda_n$. This is clearly seen in panels (b) and (c) in Fig. 12 wherein in both cases the first qubit is excited, but the kd values are different.

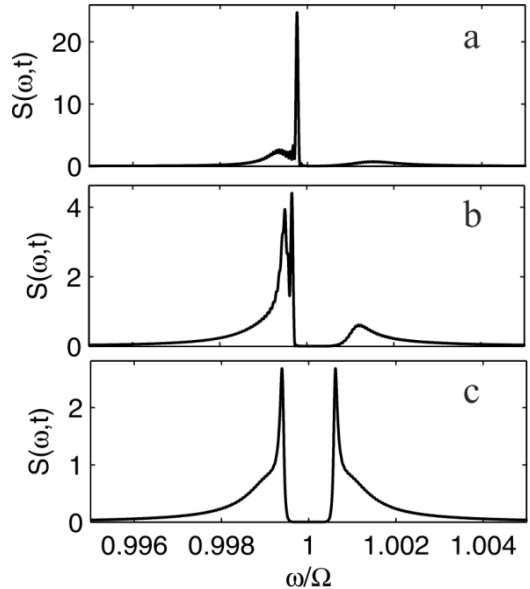

Fig. 12. Spectral density for a five-qubit system. $t=100/\Gamma$. (a) $kd=\pi/4$, central qubit is excited; (b) $kd=\pi/3$, the first qubit is excited; (c) $kd=\pi/2$, the first qubit is excited.

## VIII. Conclusion

In this paper, we study the evolution of qubits' amplitudes in a one-dimensional chain of N equidistantly spaced identical qubits where a single qubit in the chain is initially excited. For $kd=n\pi$ the amplitudes of all qubits, except for that of the excited one, are similar no matter how far the qubit under study is located from the excited qubit. In this case, a spectral line of photon emission has a Lorentzian form. It reveals a superradiant behavior with a width being equal to $N\Gamma$. However, in this case, the excitation of qubits saturates the stationary level of $1/N$. Therefore, the more is the qubit number N, the less is the amplitude of the qubit excitation. This property is a signature of the dark states which prevent the qubits from damping to zero. If kd is not equal to integer of $\pi$, the evolution of qubits' amplitudes shows damped oscillations, which are the vacuum Rabi oscillations in a multi qubit system. In this case, the amplitude of a given qubit in the chain depends on two parameters: the distance of the qubit from the excited one and the number of qubits in the chain. If N is fixed, the qubit's amplitude decreases as the distance from the excited qubit increases. If the distance of a given qubit from an excited one is fixed, the amplitude of a qubit under study decreases as N increases.

Throughout the paper, it was assumed that all qubits are identical for their frequency and the rate of spontaneous emission. Moreover, the distance between neighbor qubits was assumed to be equal. From this point, our system is similar to a chain of real two-level atoms. However, there are important differences between real atoms and superconducting qubits (artificial atoms). First, a distance between neighbors qubits is, in principle, different due to the spreading of technological parameters. Second, every superconducting qubit can be individually addressed, so that we can tune its frequency to a desirable value. Third, a distance between neighbors' qubits can purposely be made different. Therefore, it would be interesting to study how qubits' evolution could be altered by the modifications mentioned above. Undoubtedly, these issues are very important and are worthy of further investigation. We believe that the results obtained in this paper could be useful for protocols of qubit control and readout technique in superconducting circuits.

**Acknowledgments**

The work was supported by the Ministry of Education and Science of Russian Federation under the Project No. FSUN-2020-0004.